\documentclass[onecollarge,natbib]{svjour2}
\bibpunct{[}{]}{;}{n}{}{,} 
\smartqed  
\usepackage{graphicx}
\usepackage[normalem]{ulem}
 \usepackage{amsmath}
\usepackage{amssymb}
\usepackage{mathtools}
\usepackage{color}
\usepackage{xcolor}
\usepackage{lineno}
\usepackage{hyperref}
\usepackage{relsize}
\usepackage{mathrsfs}
\usepackage{bm}
\usepackage{subfigure}
\usepackage{cancel}
\usepackage{soul}

\colorlet{darkgreen}{green!50!black}
\colorlet{brightyellow}{yellow!75!red}
\colorlet{orange}{red!50!yellow}
\colorlet{darkblue}{blue!60!black}
\colorlet{darkred}{red!80!black}

\newcommand{\dd} {{\mathrm{d}}}
\newcommand{\ket}[1] {{\left.|#1\right>}}

\newcommand{\half}[1][1] {{\mathsmaller{\frac{#1}{2}}}}
\journalname{Few-Body Systems}
\begin{document}
 
\title{Non-Perturbative Calculation of the Scalar Yukawa Theory in Four-Body Truncation
       \thanks{Presented at LightCone 2014, Raleigh, North Carolina.}
}

\titlerunning{LF Scalar Yukawa}        

\author{Yang~Li         \and
        V.~A.~Karmanov  \and 
        P.~Maris        \and 
        J.~P.~Vary
}

\authorrunning{Li et al.} 

\institute{Y.~Li \and P.~Maris \and J.~P.~Vary \at
              Department of Physics and Astronomy, Iowa State University, Ames, IA 50011, USA \\
              \email{leeyoung@iastate.edu} 
\and
           V.~A.~Karmanov \at
           Lebedev Physical Institute, Leninsky Prospekt 53, 119991 Moscow, Russia \\
}

\date{\today}

\maketitle

\begin{abstract}
The quenched scalar Yukawa theory is solved in the light-front Tamm-Dancoff approach
including up to four constituents (one scalar nucleon, three scalar pions). The Fock sector dependent 
renormalization is implemented.
By studying the Fock sector norms, we find that the lowest two Fock sectors dominate 
the state even in the large-coupling region. 
The one-body sector shows convergence with respect to the Fock sector truncation.
However, the four-body norm exceeds the three-body norm at the coupling 
$\alpha \approx 1.7$.
\end{abstract}


\section{Introduction}\label{introduction}

Light-Front (LF) Hamiltonian theory is a natural \textit{ab initio} framework to study the 
structure of hadrons and strong interaction physics (see, e.g., Ref.~\cite{Bakker2013.165} and  
references therein). One of 
the challenges of any non-perturbative quantum field theory is to develop non-perturbative renormalization
schemes \cite{Perry1991.4051,Glazek1993.5863,Paston1997.1117,Grange2010.025012}.
In recent years, the Fock sector dependent renormalization (FSDR) has emerged as a promising 
systematic non-perturbative renormalization scheme and has been successfully applied to various 
field theory models \cite{Hiller1998.016006,Karmanov2008.085028,Karmanov2010.056010,Karmanov2012.085006}. 
The idea of using sector dependent counterterms in LF dynamics was first introduced, together with the LF 
Fock sector truncation (also known as the LF Tamm-Dancoff),
by Perry \textit{et al.} in Ref.~\cite{Perry1990.2959}. The authors argued that sector dependence is needed
to compensate for the non-localities introduced by the Fock sector truncation.
The use of sector dependent counterterms, based on the analysis in the explicitly covariant LF dynamics (CLFD,
see Ref.~\cite{Carbonell1998.215} for a review), 
ensures the exact cancellation of sub-divergences appearing in the Fock sectors \cite{Karmanov2008.085028}. 

Here we apply the LF Hamiltonian method with FSDR to the scalar
Yukawa theory. The theory can be used to model the pion mediated nucleon-nucleon
interaction. The Lagrangian of the theory reads,
\begin{equation} \label{eq 1}
 \mathscr{L} =
   \partial_\mu N^\dagger \partial^\mu N - m^2 |N|^2
+ \half \partial_\mu \pi \partial^\mu \pi - \half \mu^2 \pi^2 
+ g_0 |N|^2 \pi + \delta m^2 |N|^2,
\end{equation}
where $g_0$ is the bare coupling and $\delta m^2$ is the mass counterterm of the field $N(x)$. 
The physical coupling is denoted as $g$. It is convenient to introduce a dimensionless coupling 
$\alpha = \frac{g^2}{16\pi m^2}$.
We use a Pauli-Villars particle with mass $\mu_1$ to regularize the UV divergences in the theory
 \cite{Brodsky2001.114023}.
For brevity, we will refer to the fundamental degrees of freedom (d.o.f.) $N(x)$ and $\pi(x)$ as 
nucleon and pion respectively. 
The scalar cubic interaction is known to exhibit a vacuum instability \cite{Baym1960.886}. 
Following Ref.~\cite{Franz2001.076008}, we adopt the ``quenched approximation'', i.e., 
we exclude the anti-nucleon d.o.f.
Previously, the Yukawa theory with FSDR has been solved up through the three-body truncation in LF dynamics
\cite{Karmanov2008.085028,Karmanov2010.056010, Karmanov2012.085006}, 
where the renormalization has already become non-trivial. Here we extend the 
non-perturbative calculation to the quenched four-body truncation, which further demonstrates the scalability 
of the FSDR scheme. Furthermore, by comparing with lower Fock sector truncations, we investigate the convergence 
of the Fock sector expansion.

In Sect.~\ref{sec 2}, we introduce our formalism. A set of coupled integral equations will be derived. 
Numerical results are presented in Sect.~\ref{sec 3}. We summarize in Sect.~\ref{sec 4}.

\section{Formalism} \label{sec 2}

In LF dynamics, the physical states can be obtained from the eigenvalue equation,
\begin{equation}\label{schroedinger}
P^+ \hat{P}^- \ket{\psi(p)} = M^2\ket{\psi(p)},
\end{equation}
where $P^+$ is the longitudinal momentum of the system and $\hat P^-$ is the LF Hamiltonian operator. 
Thanks to the transverse boost invariance in LF dynamics, we have taken the total transverse momentum $\bm P = 0$ without the loss of generality.
In the Fock space, the state vector can be represented as,
\begin{equation}
 \ket{\psi(p)} = \sum_n \int D_n \; \psi_n(\bm k_1, x_1,\cdots \bm k_n,x_n;p)
 \ket{\bm k_1, x_1,\cdots \bm k_n,x_n},
\end{equation}
where $\psi_n(\bm k_1, x_1,\cdots \bm k_n,x_n;p) \equiv \left<\bm k_1, x_1,\cdots \bm k_n,x_n|\psi(p)\right>$ is the light-front wave function (LFWF) 
that depends on the longitudinal momentum fractions $x_i \equiv \frac{k_i^+}{P^+}$ and the relative transverse 
momenta $\bm k_i$, and $D_n$ stands for
$
2 (2\pi)^3 \delta^{(2)} (\bm k_1+\cdots \bm k_n) \delta (x_1+\cdots x_n -1) 
\prod_{i=1}^n\frac{\mathrm d^2 k_i \mathrm d x_i}{(2\pi)^3 2x_i}.
$
The LFWFs are normalized to unity, $ \sum_n I_n = 1$, where 
\begin{equation}\label{normalization factors}
I_n = \frac{1}{(n-1)!}\int D_n \left|\psi_n(\bm k_1,x_1,\cdots \bm k_n,x_n;p)\right|^2 
\end{equation}
is the probability that the system appears in the $n$-body Fock sector. 
Note that $\psi_1 = \sqrt{I_1}$ is a constant. 
It is convenient to define $s_n \equiv (k_1+k_2+\cdots k_n)^2 = \sum_{i=1}^n \frac{\bm k_i^2+m_i^2}{x_i}$ and to work with the vertex functions 
$\Gamma_n(\bm k_1,x_1,\cdots,\bm k_{n-1}, x_{n-1};p^2) \equiv (s_n - p^2) \psi_n (\bm k_1,x_1,\cdots \bm k_n,x_n;p)$. For simplicity we will omit the 
dependence on $p^2$ in $\Gamma_n$ for the ground state $p^2 = m^2$.

\begin{figure}[b]
 \centering 
  \includegraphics[width=0.8\textwidth]{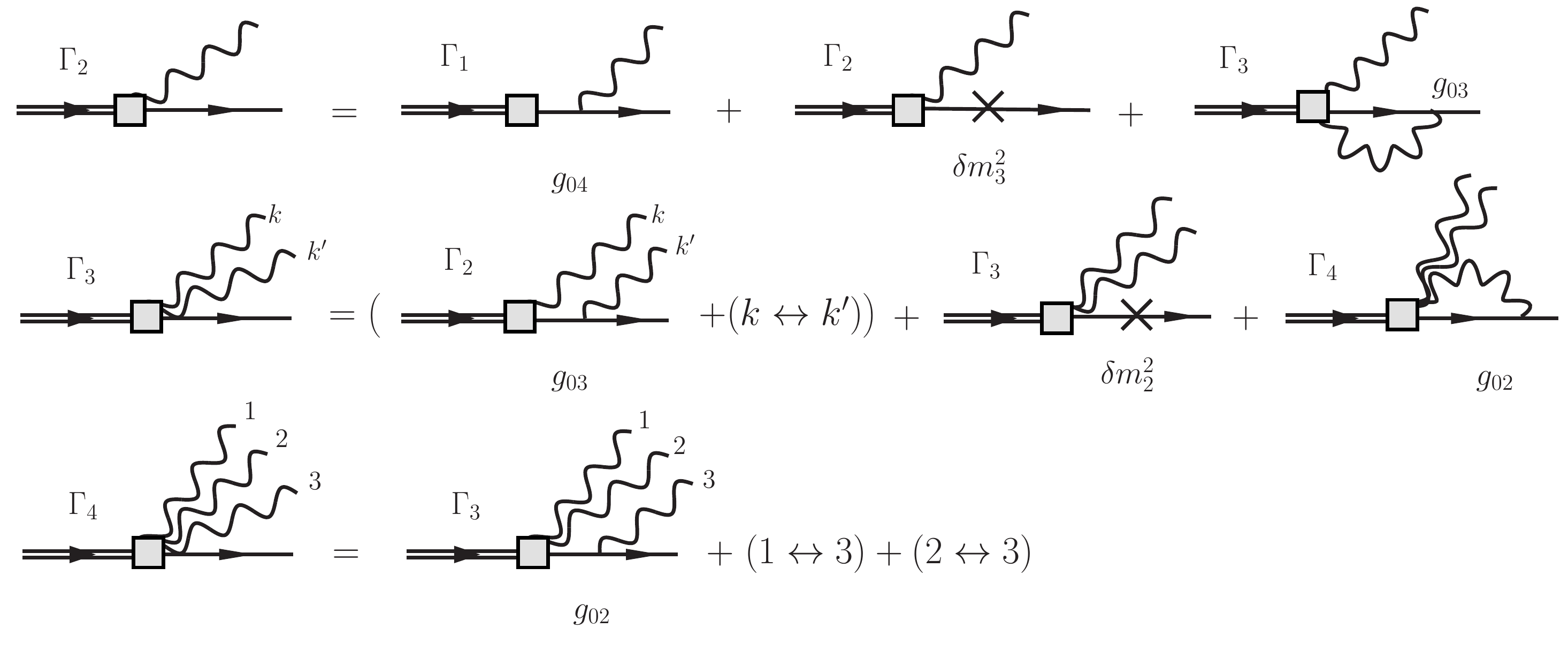}
 \caption{The diagrammatic representation of the system of equations in the four-body truncation}
 \label{fig:N4}
\end{figure}
Truncating Eq.~(\ref{schroedinger}) to the four-body Fock sector gives the system of equations shown in Fig.~\ref{fig:N4}.
Following the LF graphical rules (see, e.g., Ref. \cite{Carbonell1998.215}), the set of equations (after substituting
$\Gamma_4$ into $\Gamma_3$) read,
\begin{subequations}\label{four-body eigenvalue equations}
\begin{align}
  & \label{eq13a}
    \Gamma^j_2(\bm k, x) = g_{04} \psi_1 +  \frac{\delta m_3^2\Gamma^j_2(\bm k,x)}{(1-x)(s_2-m^2)}
 + \sum_{j'=0}^1(-1)^{j'} \!\!\int \frac{\dd^2 k'}{(2\pi)^3}\int_0^{1-x} \!\!\!\!\!\!
 \frac{\dd x' \;g_{03}(\xi')}{2x'(1-x-x')} 
   \frac{\Gamma_3^{jj'}(\bm k,x,\bm k',x')}{s_3 - m^2}, \\ 
  & \label{eq13b}
   \Gamma^{jj'}_3(\bm k,x,\bm k',x') =  
  Z^{(2)}((p-k-k')^2) \bigg[ \frac{g_{03}(\xi')\Gamma^j_2(\bm k,x)}{(1-x)(s_2-m^2)} +
     g_{02}^2 \sum_{j''=0}^1(-1)^{j''} \int \frac{\dd^2 k''}{(2\pi)^3} \int_0^{1-x-x'}  \notag \\ 
& \times \frac{\dd x''}{2x''(1-x-x'')(1-x-x'-x'')} 
 \frac{\Gamma^{jj''}_3(\bm k,x,\bm k'',x'')}{(s_{kk''}-m^2)(s_4 -m^2)} \bigg] 
 + (j,\bm k,x \leftrightarrow j',\bm k',x'),
\end{align}
where 
$\xi' = \frac{x'}{1-x}$, $\xi = \frac{x}{1-x'}$, 
$s_2 = \frac{\bm k^2 + \mu^2_j}{x} + \frac{\bm k^2 + m^2}{1-x}$, 
$s'_2 = \frac{\bm k'^2 + \mu^2_{j'}}{x'} + \frac{\bm k'^2 + m^2}{1-x'}$,
$s_3 = \frac{\bm k^2 + \mu^2_j}{x} + \frac{\bm k'^2 + \mu^2_{j'}}{x'} 
+ \frac{(\bm k+\bm k')^2 + m^2}{1-x-x'}$, 
$s_4 = \frac{\bm k^2 + \mu^2_j}{x} + \frac{\bm k'^2 + \mu^2_{j'}}{x'} + \frac{\bm k''^2 + \mu^2_{j''}}{x''}
 + \frac{(\bm k+\bm k'+\bm k'')^2 + m^2}{1-x-x'-x''}$,
$s_{kk''}= \frac{\bm k^2 + \mu^2_j}{x} + \frac{\bm k''^2 + \mu^2_{j''}}{x''} + 
\frac{(\bm k+\bm k'')^2 + m^2}{1-x-x''}$, $(p-k-k')^2 = m^2 - (1-x-x')(s_3-m^2)$, and 
$Z^{(n)}(q^2) = \left( 1 - \frac{\Sigma^{(n)}(q^2) - \Sigma^{(n)}(m^2)}{q^2-m^2} \right)^{-1}$ 
is a generalization of the $Z$-factor (the field strength renormalization constant) coming from the $n$-body 
self-energy corrections $\Sigma^{(n)}$. 
Here $j=0,1$ refers to the physical ($\mu_0 = \mu$) and PV pion, respectively. As mentioned, the counterterms
admit Fock sector dependence. $g_{02}, g_{03}, \delta m_2^2, \delta m_3^2$
are the two- and three-body renormalization parameters that have been obtained from the two- and three-body truncation.
Note that the three-body bare coupling $g_{03}$ depends on the \textit{relative} longitudinal momentum fraction, which is a consequence
of the three-body truncation \cite{Karmanov2008.085028,Karmanov2012.085006}. 
\end{subequations} 

In order to obtain the four-body bare coupling $g_{04}$, we apply the renormalization condition 
\cite{Karmanov2008.085028} 
\begin{equation}\label{renormalization condition}
 \Gamma_2^{j=0}(\bm k^\star, x) = g\sqrt{I_1^{(3)}},
\end{equation}
where $I_1^{(3)}$ is the one-body norm found in the three-body truncation and 
$s_2^\star\equiv \frac{ {\bm k^\star}^2 + \mu^2}{x}+\frac{{\bm k^\star}^2 + m^2}{1-x} = m^2$ is on the mass shell.
\begin{figure}[b]
 \centering 
 \includegraphics[width=0.6\textwidth]{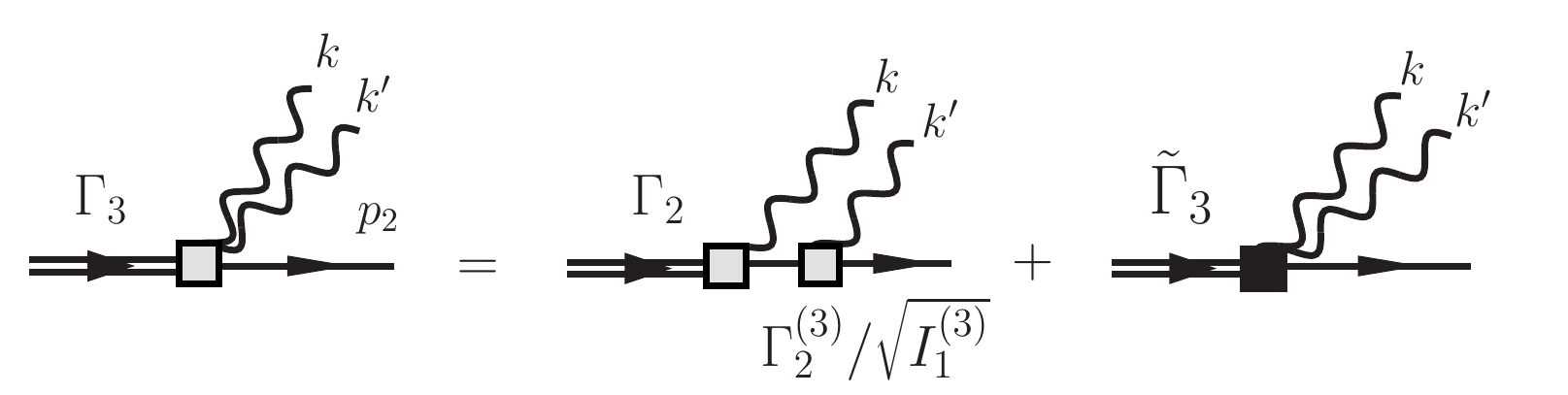}
 \caption{Split $\Gamma_3$ using the two-body vertex $\Gamma_2^{(3)}$ found in the three-body truncation}
 \label{fig:splitGamma3}
\end{figure}
However, Eq.~(\ref{eq13a}) contains a mass pole $\frac{1}{s^\star_2-m^2}$, after applying the renormalization 
condition Eq.~(\ref{renormalization condition}). 
This pole is associated with the three-body mass counterterm $\delta m_3^2$. Eq.~(\ref{eq13b}) 
contains a similar mass pole, presumably associated with the three-body loop correction.  
These poles should cancel and produce a $Z$-factor when combined.
To facilitate the analytical cancellation, we can split $\Gamma_3$ into two pieces (see Fig.~\ref{fig:splitGamma3}), 
\begin{equation}\label{split Gamma3}
  \Gamma^{0j'}_3(\bm k,x,\bm k',x') \equiv \frac{\Gamma_2^{j'(3)}(\bm q',\xi';(p-k)^2)}{\sqrt{I^{(3)}_1}}
\frac{ \Gamma^0_2(\bm k,x)}{(1-x)(s_2-m^2)} 
+ \widetilde\Gamma^{0j'}_3(\bm k,x,\bm k',x') , 
\end{equation}
where $\Gamma^{(3)}_2$ is  the two-body vertex function found in the three-body truncation, 
$\widetilde\Gamma^{0j'}_3(\bm k,x,\bm k',x')$ is a regular function when $s_2$ is taken on mass-shell and 
$(p-k)^2 = m^2 - (1-x)(s_2-m^2), \xi' = \frac{x'}{1-x}, \bm{q}' = \bm{k}' + \xi'\bm{k}$. The first part on the right-hand
side of Eq.~(\ref{split Gamma3}) produces the necessary three-body loop correction when plugged into Eq.~(\ref{eq13a}).
We substitute $\Gamma^{0j'}_3$ from Eq.~(\ref{split Gamma3}) in Eq.~(\ref{eq13a}) and then apply the renormalization 
condition Eq.~(\ref{renormalization condition}). Note that thanks to FSDR, the $Z$-factor 
still equals the one-body norm, $Z^{(3)} \equiv \left( 1 - \frac{\partial}{\partial p^2}\Sigma^{(3)}(p^2) \right)^{-1}_{p^2=m^2} = I_1^{(3)}$, even under the Fock sector truncation \cite{Karmanov2010.056010}.
Then $\Gamma_2$ becomes,
\begin{multline}\label{Gamma2}
  \Gamma^j_2(\bm k, x) = {g}/{\sqrt{I_1^{(3)}}} + \delta m_3^2 \frac{\Gamma^j_2(\bm k,x)}{(1-x)(s_2-m^2)} \\
  + \sum_{j'=0}^1(-1)^{j'} \int \frac{\dd^2 k'}{(2\pi)^3} 
 \int_0^{1-x}\frac{\dd x' \;g_{03}(\xi')}{2x'(1-x-x')}
    \left[ \frac{\Gamma_3^{jj'}(\bm k,x,\bm k',x')}{s_3 - m^2} 
- \frac{\widetilde\Gamma^{0j'}_3(\bm k^\star,x,\bm k',x')}{s_3^\star - m^2} \right],
\end{multline}
where 
$s^\star_3 = \frac{{\bm k^\star}^2 + \mu^2}{x} + \frac{\bm k'^2 + \mu^2_{j'}}{x'} 
+ \frac{(\bm k^\star+\bm k')^2 + m^2}{1-x-x'}$.
The on-mass-shell condition, after some arithmetic ${\bm k^\star}^2 = -(1-x)\mu^2-x^2m^2$,
implies that $\widetilde\Gamma_3^{0j'}(\bm k^\star,x,\bm k',x')$ is an analytic continuation of 
$\widetilde\Gamma_3^{jj'}(\bm k,x,\bm k',x')$, which can not be directly rendered in the numerical calculation. 
Therefore, we keep $\widetilde\Gamma_3^{0j'}(\bm k^\star, x,\bm k',x')$ as 
an auxiliary function that satisfies its own integral equation, 
\begin{multline}\label{tildeGamma3}
  \widetilde\Gamma^{0j'}_3(\bm k^\star,x,\bm k',x') =  
 Z^{(2)}((p-k^\star-k')^2) \Bigg[ 
   g_{02}^2 \sum_{j''=0}^1\int \frac{\dd^2 k''}{(2\pi)^3} \int_0^{1-x-x'} \!\!\! \!\!\!\!\!\!\!\! \!\!\!\!
\frac{\dd x''}{2x''(1-x-x'')(1-x-x'-x'')}   \\
\times \frac{(-1)^{j''}}{s^\star_4 -m^2} 
\bigg( \frac{\widetilde\Gamma^{0j''}_3(\bm k^\star,x,\bm k'',x'')}{s_{k^\star k''}-m^2} 
+ \frac{\Gamma^{j'j''}_3(\bm k',x',\bm k'',x'')}{s_{k'k''}-m^2} \bigg) +
\frac{g_{03}(\xi)\Gamma^{j'}_2(\bm k',x')}{(1-x')(s'_2-m^2)}  
\Bigg], 
\end{multline}
where 
$s_{k^\star k''}= \frac{{\bm k^\star}^2 + \mu^2}{x} + \frac{\bm k''^2 + \mu^2_{j''}}{x''} + \frac{(\bm k^\star+\bm k'')^2 + m^2}{1-x-x''}$, 
$s_4^\star = \frac{{\bm k^\star}^2 + \mu^2_j}{x} + \frac{\bm k'^2 + \mu^2_{j'}}{x'} + \frac{\bm k''^2 + \mu^2_{j''}}{x''}
 + \frac{(\bm k^\star+\bm k'+\bm k'')^2 + m^2}{1-x-x'-x''}$,
$s_{k'k''}= \frac{\bm k'^2 + \mu^2_j}{x'} + \frac{\bm k''^2 + \mu^2_{j''}}{x''} + 
\frac{(\bm k'+\bm k'')^2 + m^2}{1-x'-x''}$, and $(p-k^\star-k')^2 = m^2 - (1-x-x')(s^\star_3-m^2)$. 

Eqs.~(\ref{eq13b}, \ref{Gamma2}, \ref{tildeGamma3}) constitute the properly renormalized system of equations.

\section{Numerical Results} \label{sec 3}

\begin{figure}[b]
 \centering 
 \includegraphics[width=0.8\textwidth]{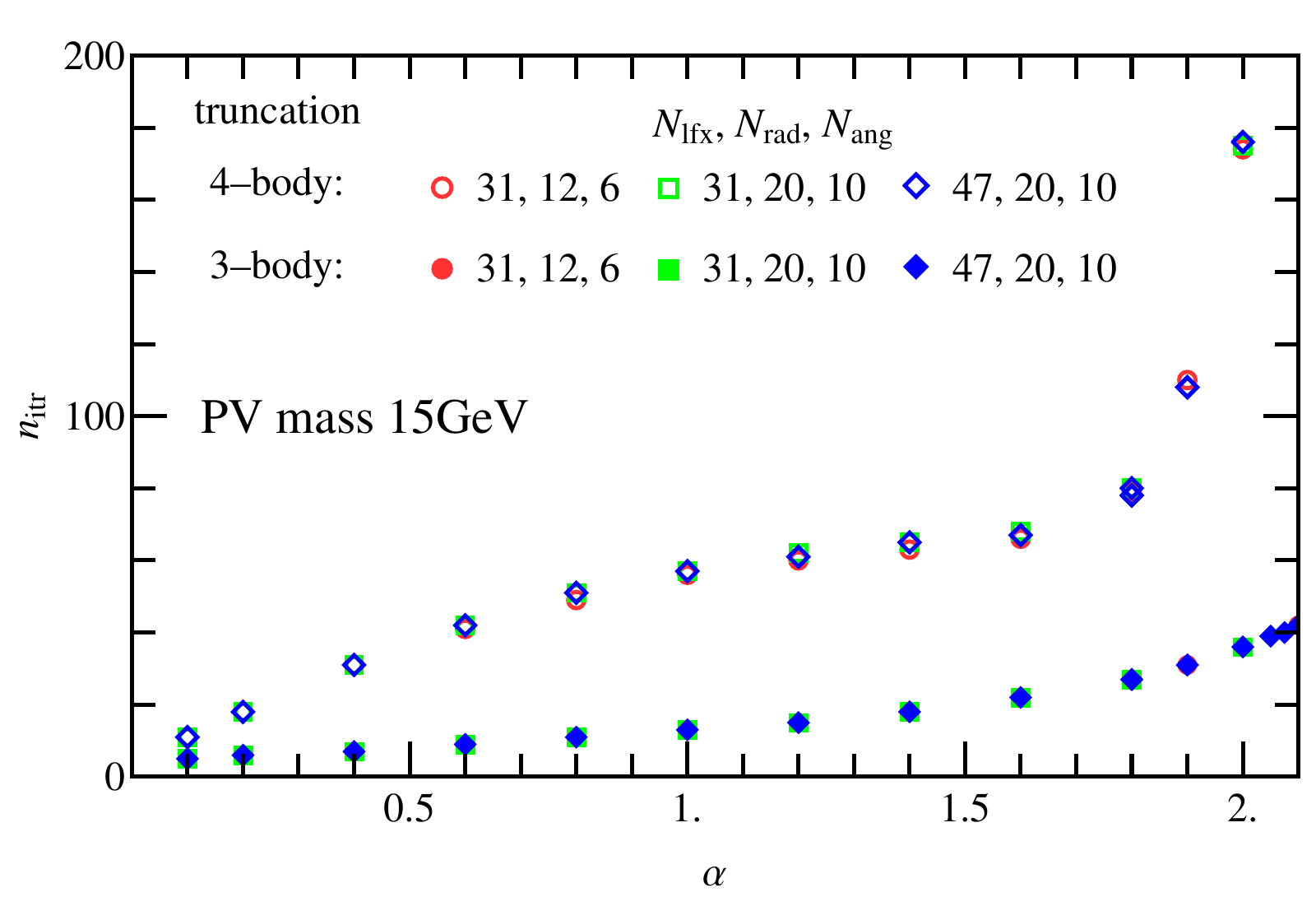}
 \caption{ The number of iterations needed to achieve absolute convergence $10^{-4}$ for different couplings. 
Numerical difficulty is detected beyond $\alpha = 1.8$
}
 \label{fig:nitr}
\end{figure}

We employ an iterative method to solve the system of equations Eqs.~(\ref{eq13b}, \ref{Gamma2}, \ref{tildeGamma3}) 
numerically.
The vertex functions are discretized on a 5-dimensional grid. 
The size of the grid is proportional to $N^2_\mathrm{lfx} N^2_\mathrm{rad} N_\mathrm{ang}$, 
where $N_\mathrm{rad}, N_\mathrm{ang}, N_\mathrm{lfx}$ are the number of grid points in the transverse radial, 
angular and the longitudinal directions, respectively. 
For the integrations we use Gauss-Legendre quadrature methods, interpolating on the external grid points as necessary. 
We use the solution of the three-body truncation as the initial guess for the vertex functions $\Gamma$ 
and then update them iteratively, until they, before and after the update, satisfy the criterion
$\max\big\{| \Gamma(\text{after}) - \Gamma({\text{before}}) |\big\} < 10^{-4}$.  
Typically, we need 50 to 100 iterations to reach convergence (see Fig.~\ref{fig:nitr}). 
\begin{figure}[tb]
 \centering 
 \includegraphics[width=0.8\textwidth]{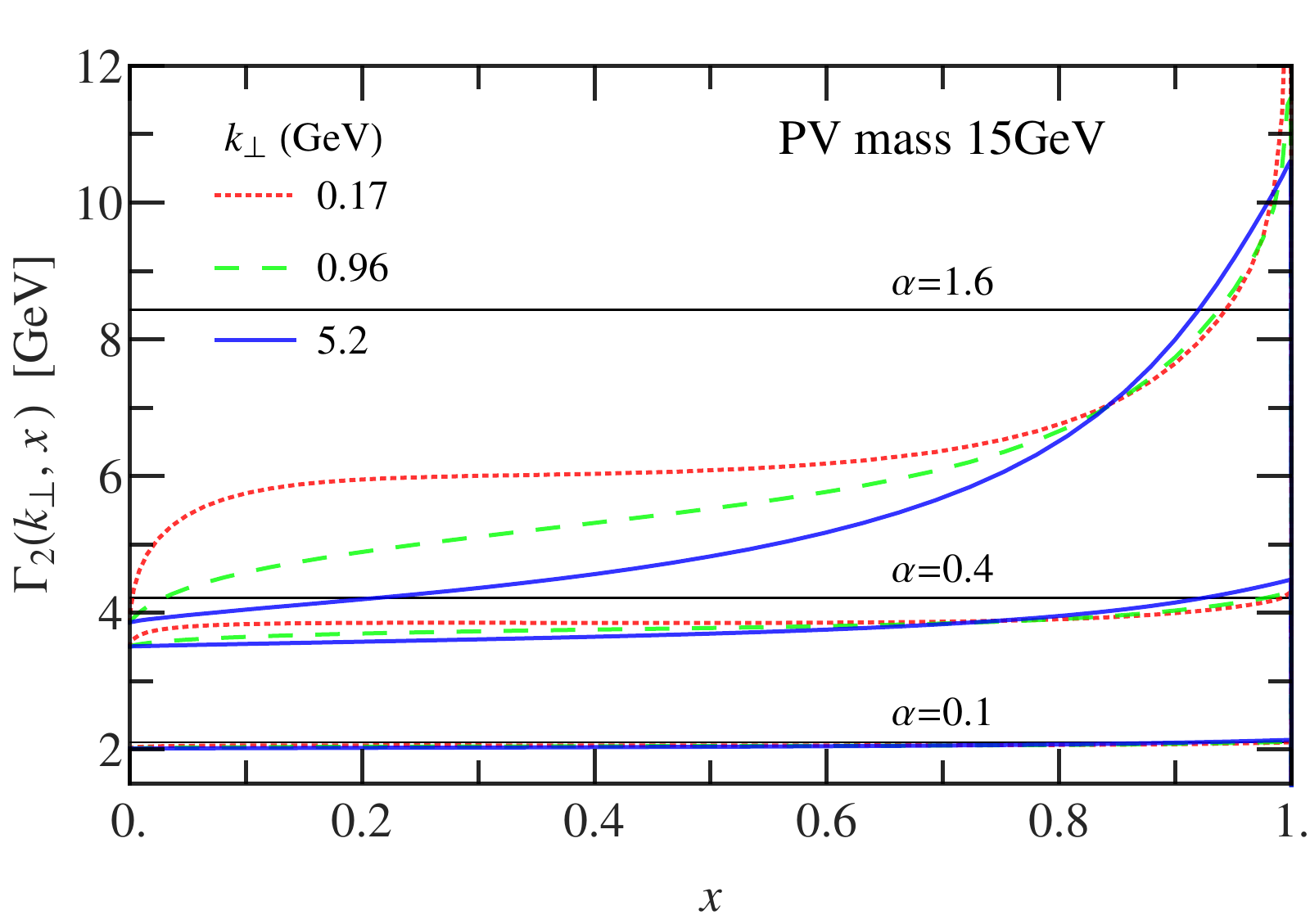}
 \includegraphics[width=0.8\textwidth]{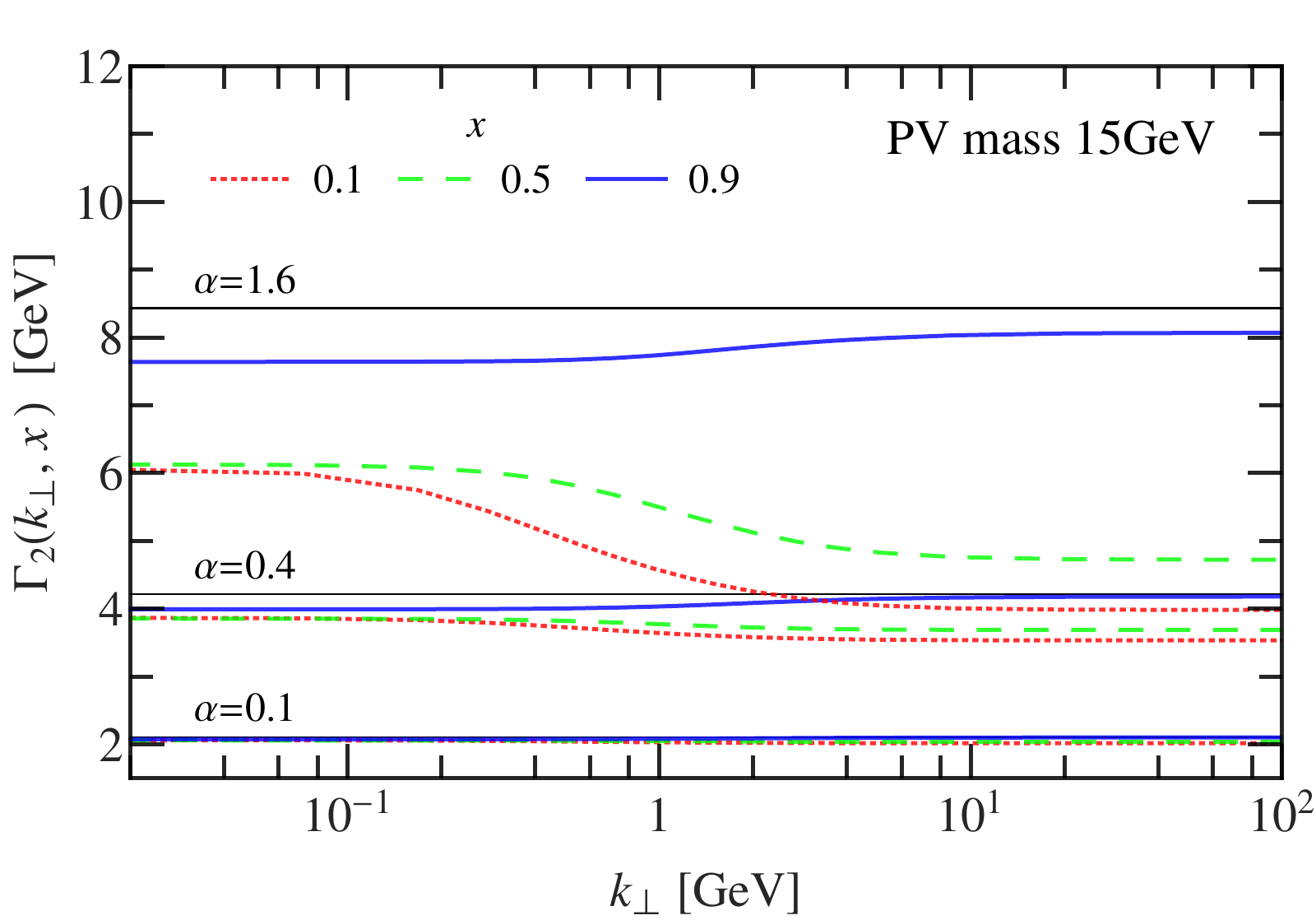}
 \caption{The two-body vertex function $\Gamma_2(k_\perp,x)$ for $\alpha = 0.1, 0.4, 1.6$,
  on a grid $N_\mathrm{rad} = 47, N_\mathrm{ang} = N_\mathrm{lfx} = 20$. The horizontal
lines are corresponding results from the one-loop LF perturbation theory
}
 \label{fig:vertexfunction}
\end{figure}
\begin{figure}[tb]
 \centering 
 \includegraphics[width=0.8\textwidth]{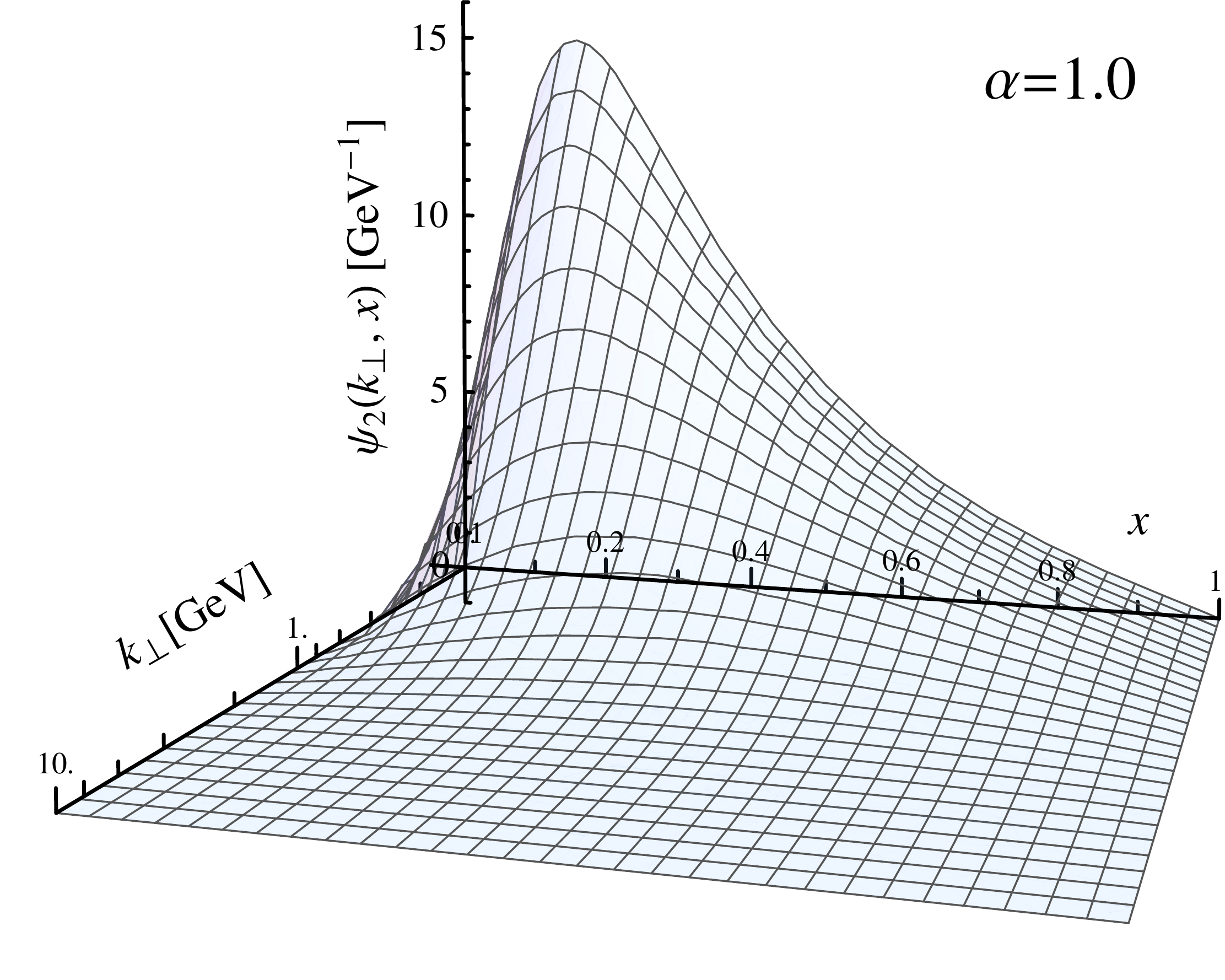}
 \caption{The two-body LFWF $\psi_2(k_\perp,x)$ for $\alpha = 1.0$.
  Grid size: $N_\mathrm{rad} = 47, N_\mathrm{ang} = N_\mathrm{lfx} = 20$}
 \label{fig:psi2in3D}
\end{figure}
The numerical calculations were carried out on the Cray XE6 Hopper at NERSC with a hybrid MPI/OpenMP code.
The system was solved at $\mu = 0.14\;\mathrm{GeV}, m = 0.94\; \mathrm{GeV}, \mu_1 = 15 \,\mathrm{GeV}$ for
various couplings. 
We have also investigated, analytically and numerically, the dependence on the PV mass as it goes to infinity.
We found very weak PV mass dependence for $\mu_1 > 10\,\mathrm{GeV}$. More details will be given in 
Ref.~\cite{Li in preparation}.
Representative two-body vertex functions are shown in Fig.~\ref{fig:vertexfunction}. In the one-loop 
LF perturbation theory, the dressed vertex is merely a constant $g$, shown as horizontal lines in
Fig.~\ref{fig:vertexfunction}. At small coupling, our two-body vertex function $\Gamma_2$ reproduces the perturbation 
result.
From the vertex functions, we can obtain the LFWFs. Fig.~\ref{fig:psi2in3D} shows the two-body LFWF for $\alpha=1.0$.

\begin{figure}[ht]
 \centering
\includegraphics[width=0.8\textwidth]{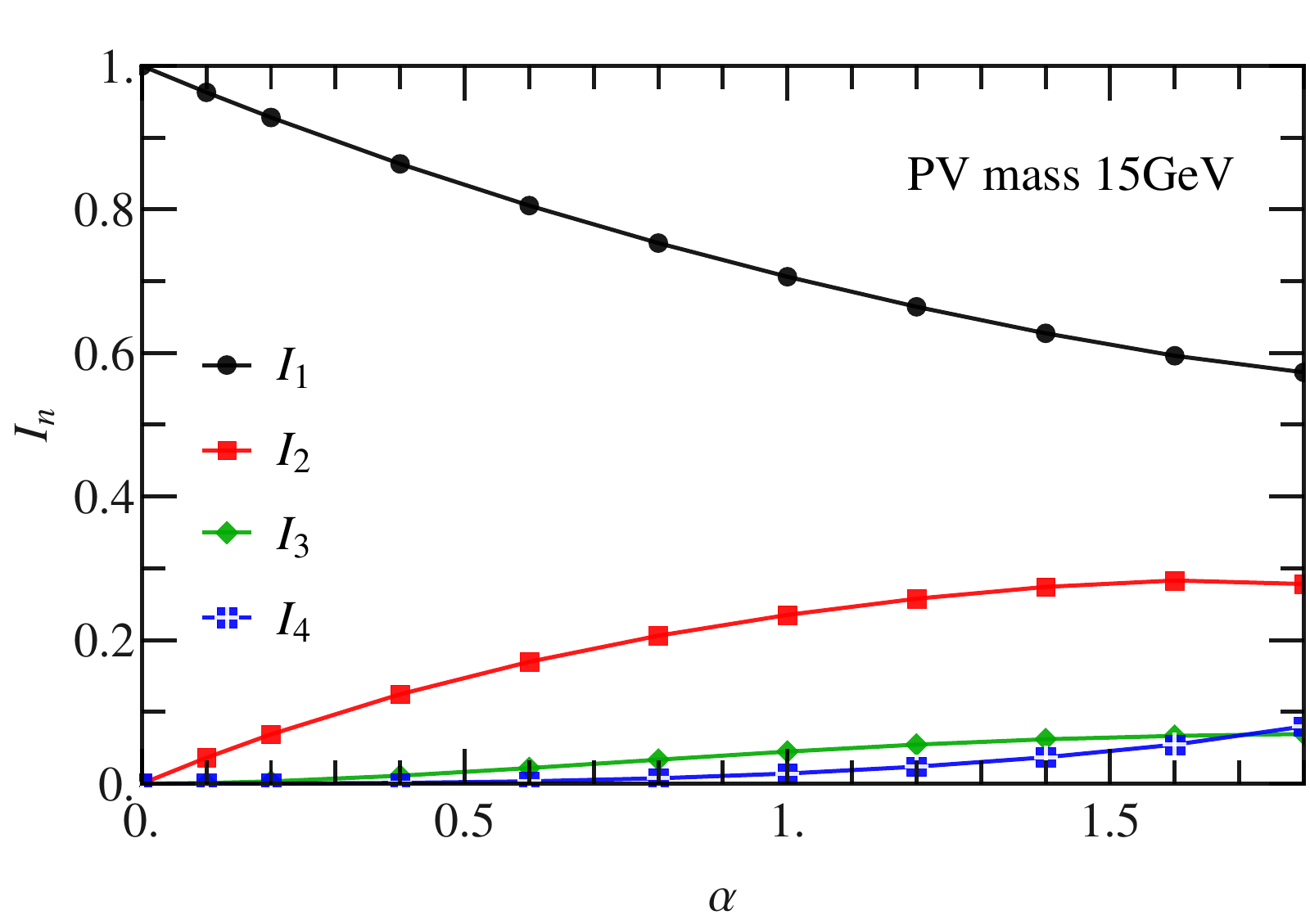}\label{fig:FockComponents_N4_a} 
\includegraphics[width=0.8\textwidth]{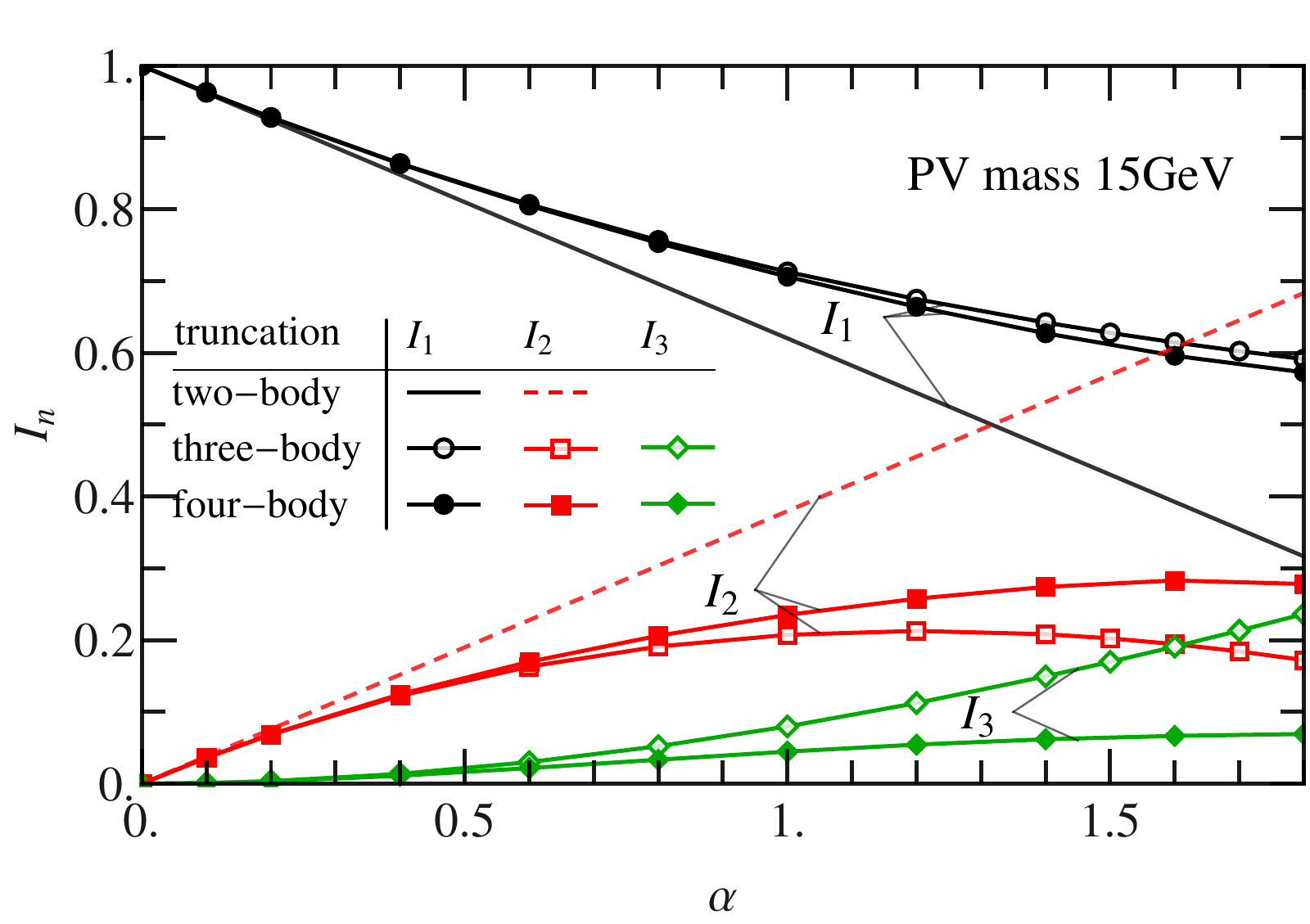} \label{fig:saturation_of_all}
 \caption{
Top panel: the $n$-body Fock sectors norm $I_{n}$ as a function of the coupling $\alpha$ in the four-body truncation;
Bottom panel: $I_n$ as a function of the coupling $\alpha$ in the two-, three- and four-body truncations
}
 \label{fig:FockComponents_N4}
\end{figure}

The top panel of Fig.~\ref{fig:FockComponents_N4} plots the norm of the Fock sectors (see Eq.~(\ref{normalization factors})) 
in the four-body truncation as a function of the coupling up to $\alpha = 1.8$. 
The contributions to the state vector show a hierarchy of Fock sectors (i.e. $I_n < I_{n-1}$) up to $\alpha \approx 1.7$. 
Beyond that $I_4$ becomes larger than $I_3$. Nevertheless, the lowest two sectors $\ket{N}+\ket{N\pi}$ are observed 
to dominate the Fock space up to $\alpha = 1.8$, where these two sectors constitute about $85\%$ of the full norm. 
The bottom panel of Fig.~\ref{fig:FockComponents_N4} compares the four-body truncation with the two- and three-body Fock sector 
truncations. The one- and two-body Fock sector norm $I_1$ and $I_2$ show trend of convergence as the number of 
Fock sector constituents increase, especially below $\alpha \approx 1.0$. 
Note that the one-body sector $I_1$ changes little from the three-body truncation to the four-body truncation, 
even at large coupling. $I_1$ is closely related to observables, in particular, the electromagnetic form factor
$F(Q^2 \to \infty) \propto I_1$, and the parton distribution function of the nucleon $f(x) = I_1\delta(x-1) + \cdots$.

The convergence of the Fock sector truncation breaks down once the contribution of the highest Fock sector supersedes 
the contribution from the lower Fock sectors.
In terms of the Fock sector norms,
the breakdown of the two-body truncation ($I_2>I_1$) happens at $\alpha \approx 1.3$; the three-body truncation
($I_3>I_2$) at $\alpha \approx 1.6$; the four-body truncation ($I_4>I_3$) at $\alpha \approx 1.7$.
This suggests that the non-perturbative results can be systematically improved by including more Fock sectors. 
%
\section{Summary} \label{sec 4}

We study the scalar Yukawa theory in the non-perturbative region (up to $\alpha = 1.8$) within the four-body 
truncation (one nucleon and three pions). The theory is quantized on the light front and a Fock sector dependent renormalization is 
implemented. The system of equations for the one-nucleon sector is derived and solved numerically.  
The investigation of the Fock sector norms largely supports the hierarchy picture up to about $\alpha \approx 1.7$: 
the system is dominated by the lowest Fock sectors and at a fixed coupling the Fock sector expansion shows convergence as 
the number of constituents increases. 

 The success of solving a four-body truncated quantum field theory in the non-perturbative region gives us further 
confidence for the use of FSDR together with LFTD as a general \textit{ab initio} approach.
The study of the higher Fock sector expansion in more realistic field theories such as the Yukawa model, QED and QCD models 
is in progress.
Solving the one-nucleon sector also provides the first step to study the bound-state problem in the two-nucleon sector,
 which has been extensively studied in various approaches \cite{Hiller1993.4647,Savkli1999.055210,Hwang2004.413,Ji2012.054011}.
However not all these approaches are from first principles, nor do they include a systematic non-perturbative 
renormalization. The investigation of this problem in the LFTD approach is also under consideration.
%

\begin{acknowledgements}
We are indebted to A.V.~Smirnov for kindly providing us some numerical benchmark results for three-body truncation.
We wish to thank J.~Carbonell, J.-F.~Mathiot and  Xingbo Zhao for valuable discussions. One of us (V.A.K.) is 
sincerely grateful to the Nuclear Theory Group at Iowa State University for kind hospitality during his visits.
This work was supported in part by the Department of Energy under Grant Nos. DE-FG02-87ER40371 
and DESC0008485 (SciDAC-3/NUCLEI) and by the National Science Foundation under Grant No
PHY-0904782.  
Computational resources were provided by the National Energy Research Supercomputer Center (NERSC), 
which is supported by the Office of Science of the U.S. Department of Energy under Contract 
No. DE-AC02-05CH11231.

\end{acknowledgements}


\end{document}